\documentstyle[twocolumn,aps]{revtex}
\date{\today}
\input psfig.tex
\begin{document}
\draft
\twocolumn[\hsize\textwidth\columnwidth\hsize\csname 
@twocolumnfalse\endcsname
\title{
Non--Zero Fermi Level Density of States for a Disordered d-Wave Superconductor 
in Two Dimensions}
\author{ K. Ziegler$^a$, M.H. Hettler$^b$, and P.J. Hirschfeld$^b$}
\address{$^a$Max-Planck-Institut f\"ur Physik Komplexer Systeme,
Au\ss enstelle Stuttgart, Postfach 800665, D-70506 Stuttgart, Germany\\
$^b$ Dept. of Physics, Univ. of Florida, Gainesville, FL 32611, USA.}
\maketitle 
\begin{abstract}
It has been known for some time that, in three dimensions, arbitrarily weak
disorder in unconventional superconductors with line nodes 
gives rise to a nonzero residual density of zero-
energy quasiparticle states $N(0)$, leading 
to characteristic low-temperature thermodynamic
properties similar to those observed in cuprate and 
heavy-fermion systems.  In a strictly two-dimensional 
model possibly appropriate for the cuprates,  
it has been argued that $N(0)$ vanishes, however.  
We perform exact calculations for  d- and extended s-wave superconductors 
with Lorentzian disorder and  similar models with 
continuous disorder
distribution, and show that in these cases the residual density of states 
is nonzero even in two dimensions. We discuss the reasons for this
discrepancy, and the implications of our result for the cuprates.
\end{abstract}
]
\narrowtext
{\it Introduction.}  A good deal of evidence \cite{AGL}
has accumulated
recently suggesting that the order parameter in the cuprate
superconductors vanishes linearly at lines on the Fermi surface.   
Frequently these experiments have been interpreted in terms of a 
$d_{x^2-y^2}$ pairing state, but states of ``extended-s" symmetry
with nodes have also been considered.  One of the most interesting
consequences of such nodes in three spatial dimensions is the creation
of a nonzero density of zero-energy quasiparticle states $N(0)$ for
infinitesimal disorder.\cite{GorkovKalugin,UedaRice}  Such a residual 
density of states (DOS) is of course reflected in many experimental
observables, and may be shown \cite{gross}
to lead in particular to a $T^2$ term
in the London penetration depth $\lambda(T)-\lambda(0)$, and more 
generally to low-temperature thermodynamic and transport properties
characteristic of a normal Fermi system with strongly reduced 
DOS.  Systematic $Zn$-doping and electron damage 
experiments have been found to lead to precisely these types of temperature
dependences in $YBCO$ single crystals\cite{Ishida},
and in certain cases quantitative
fits \cite{HPSPRB} have been obtained to the ``dirty d-wave" model, 
in which the
effects of potential scatterers on a 2D d-wave superconductor
are calculated using a t-matrix approximation \cite{ph1,srink}
assuming large 
electronic phase shifts. \\

Recently, Nersesyan
et al. \cite{Nersetal} 
have questioned the accuracy of the t-matrix approximation
when applied to a strictly 2D disordered d-wave system,
pointing out that in 2D logarithmic divergences in multisite 
scattering processes, some of which are neglected in the t-matrix approach, 
prevent a 
well-controlled expansion in impurity concentration.
These authors avoid perturbation theory by using bosonization together with 
the replica trick, and predicted a power law DOS $N(E)
\sim |E|^\alpha$, $\alpha\simeq 1/7$, for sufficiently
small energy $E$ and disorder, rather than the analytic behavior
$N(\omega)\sim const+a E^2$ expected in 3D.
They also argued that a non-zero DOS at $E=0$, a quantity indicating
spontaneous symmetry breaking, may not occur because of the Mermin-Wagner
theorem \cite{MW}.\\

Although the physical systems in question are in reality
highly anisotropic 3D systems,
the possibility of a 2D-3D crossover at low temperatures could conceivably
invalidate some of the results of the usual ``dirty d-wave" approach.
This would render the description of the low-temperature transport properties
of the cuprate superconductors considerably more complicated even
were the order parameter to correspond to the very simple 2D
$d_{x^2-y^2}$ form usually assumed.  \\

It is therefore of considerable importance to check 
the results of Ref. \cite{Nersetal} 
by other methods. In this paper we show that for certain
types of disorder, {\it exact} results can be obtained for the DOS of
strictly 2D disordered superconductors.  
We show that for {\it any} disorder diagonal in position and
particle-hole space, the DOS of a classic isotropic s-wave
superconductor has a rigorous threshold at the (unrenormalized)
gap edge $\Delta$, as expected from Anderson's theorem \cite{Anderson}.
Within the same general method, we show that the residual DOS $N(0)$
of a superconductor
with line nodes (e.g. d- or extended s-wave) is nonzero for arbitrarily
small disorder, in disagreement with Ref. \cite{Nersetal}. 
We believe that the DOS in a disordered system 
is not an order parameter which belongs to the class
of order parameters covered by the Mermin--Wagner theorem.
This is supported by the fact that a non-zero DOS occurs also
in other tight-binding models (e.g., model for two-dimensional Anderson
localization\cite{RL}), which are described
by a field theory with continuous symmetry.
\vskip .2cm
As exact results are only obtainable for Lorentzian
disorder, we discuss ways \cite{ziegler2} of obtaining information 
on the effects of 
other distributions, including models where the randomness has 
a compact domain.  Finally, we compare our results to those arising
from alternative methods, and comment on possible origins of the current
disagreement.\\

{\it Density of states.}  Here, we introduce a general method of calculating 
exactly the DOS of a superconductor 
for certain types of disorder, motivated
by the analysis of Dirac fermions in 2D. \cite{ziegler1} The BCS  
Hamiltonian  is given by
\begin{equation}
H=(-\nabla^2-\mu)\sigma_3+\Delta\sigma_1,
\label{ham}
\end{equation}
which describes quasiparticles in the presence of the
spin singlet order parameter $\Delta$.  The $\sigma_i$ are the Pauli matrices
in particle-hole space. The disorder is modeled by taking $\mu=\mu_x$ as
a random variable distributed according to a probability distribution
$P(\mu_x)$.\\

The kinetic energy operator $-\nabla^2$ is taken to
act as $\nabla^2\Psi(x)=\Psi(x+2e_1)+\Psi(x-2e_1)+\Psi(x+2e_2)
+\Psi(x-2e_2)$ on  a function $\Psi(x)$ of the 
sites $x$ of a 2D square lattice spanned by the unit
vectors $e_1$ and $e_2$.  Note this function involves displacements
of two lattice sites rather than one, as would be the case 
in the simplest  tight-binding representation of the lattice kinetic energy.
For a system of fermions in the thermodynamic limit, the bare kinetic energy
will then have a band representation quite similar to 
the usual tight-binding form, with no particular distinguishing features near
the Fermi level.  The reason for this choice will become clear below. It obeys,
of course, the same global continuous symmetries discussed for the model in
Ref. \cite{Nersetal}.
The bilocal lattice operator ${\hat \Delta}\equiv\Delta_{x,x^\prime}$ is taken
to act as a c-number in the isotropic s-wave case,
${\hat\Delta}\Psi(x)=\Delta\Psi(x)$,  whereas to study
extended pairing we define ${\hat \Delta}^{s\atop d}\Psi(x)=\Delta^{s\atop d}
[\Psi(x+e_1)+\Psi(x-e_1)\pm\Psi(x+e_2)\pm\Psi(x-e_2)]$.\\

We consider the single-particle Matsubara Green function 
defined as 
$G(iE)=(iE\sigma_0 - H)^{-1}$. We are  primarily interested in
calculating the DOS 
$N (E)\equiv -\frac{1}{\pi} {\rm Im} \sum_{\vec{k}} 
\langle G_{1\,1}({\vec k},iE \rightarrow E + i\epsilon)\rangle$
where $\langle ...\rangle$ denotes the disorder average.  
The problem now is how to perform this disorder average over the 
probability measure $P(\mu_x)d\mu_x$ of the random variable $\mu_x$. Exact
results for the disorder-averaged propagator in
noninteracting systems can frequently be obtained for Lorentzian disorder,
$P(\mu_x)d\mu_x=(\gamma/\pi)[(\mu_x-\mu_0)^2+\gamma^2]^{-1}d\mu_x$, 
by exploiting the simple pole
structure of $P(\mu_x)$ in the complex $\mu_x$ plane. $\mu_0$ is
the chemical potential of the averaged system.  For convenience, we 
set $\mu_0 = 0$.
 The averaged Green
function is $\langle G(iE)\rangle\equiv \int \prod_x d\mu_x
 P(\mu_x) G(iE;\mu_x)$, which may
then be trivially evaluated {\it if} $G$ can be shown to be analytic in
either the upper or lower half-plane.\\

In a superconductor, the Green
function depends on the random variable $\mu_x$ via $\mu_x\pm iE$,
as a consequence of the particle-hole structure. Therefore, the averaging of
$G$ with respect to Lorentzian disorder is not trivially possible.
However, we will show below that it is possible to reformulate the problem 
so that $G$ consists of terms which  are analytic in one of the half planes.
This allows us to perform the averaging of the Green function for 
Lorentzian disorder. \\

{\it Isotropic s-wave superconductor.}  We  first assume a homogeneous s-wave order
parameter, neglecting the response of the superconducting condensate to the
random potential.  The Matsubara Green function may be written
$G(iE)=-(iE\sigma_0 + H)(E^2 + H^2)^{-1} $
where we note that $H^2 = (-\nabla^2-\mu)^2\sigma_0+\Delta^2\sigma_0$
since in the isotropic s-wave case,  $(-\nabla^2-\mu)\sigma_3$ anticommutes 
with $\Delta\sigma_1$ even for random $\mu$ due to the locality of the 
order parameter.\\

The expression $ H^2+E^2$ is proportional to the unit
matrix; as a consequence, the Green  function can be written in the simple
form
\begin{eqnarray}
& G(iE) =  -{iE\sigma_0 + H\over2i\sqrt{\Delta^2+E^2}} &
\left[(-\nabla^2-\mu-i\sqrt{\Delta^2+E^2})^{-1} \right. \nonumber\\
& & \left. -(-\nabla^2-\mu+i\sqrt{\Delta^2+E^2})^{-1}\right]\sigma_0
\label{gswave1}
\end{eqnarray}
It is straightforward to show that the imaginary part of this expression 
(after analytic continuation, $iE\rightarrow E + i \epsilon$)
for {\it any given
configuration of impurities} is vanishing for $|E| < \Delta$.
Therefore, the DOS shows  a gap of size $\Delta$ 
{\it independent} of the distribution function $P(\mu)$.
Thus, our model reproduces the famous Anderson
theorem\cite{Anderson} which states that the thermodynamics of an isotropic
s-wave superconductor are not affected by diagonal, nonmagnetic disorder. The
situation is different if the order parameter itself is random 
\cite{ziegler2}.\\

{\it  d- and extended-s symmetry superconductors.}  The second 
class of examples includes the d-wave and extended-s 
"bond" order parameters ${\hat \Delta}^{s\atop d}$ defined above. 
The corresponding pure systems in momentum space fulfill the 
condition $\sum_k\Delta_k=0$, so that nonmagnetic disorder must 
cause significant pair breaking \cite{GorkovKalugin}.
The behavior of the imaginary part of the
Green's function can be studied using a method analogous to that used for
the s-wave case. However, the main difference is that the nonlocal 
order parameter term ${\hat \Delta}^{s  \atop d}\sigma_1$ does not anticommute 
with $(-\nabla^2-\mu)\sigma_3$ anymore
if $\mu$ is random.\\

This requires a different type of transformation. We
introduce a diagonal matrix (or staggered field)
$D_{x,x'}=(-1)^{x_1+x_2}\delta_{x,x'}$ (note $D^2$ is the unit matrix).
Now we may write
\begin{eqnarray}
& H^2=HD\sigma_3^2DH= & [(-\nabla^2-\mu)D\sigma_0-i{\hat \Delta}^{s  \atop d}
D\sigma_2] \nonumber \\
& &  [D(-\nabla^2-\mu)\sigma_0+iD{\hat \Delta}^{s  \atop d}\sigma_2]
\label{6}
\end{eqnarray}
Because $D$ commutes with $-\nabla^2$ (as defined above) and $\mu$, but
anticommutes with the order parameter ${\hat \Delta}^{s  \atop d}$, 
we have simply $H^2={\tilde H}^2$, with $\tilde H\equiv 
(-\nabla^2-\mu)D\sigma_0-i{\hat \Delta}^{s  \atop d} 
D\sigma_2$. Therefore, the quantity $H^2+E^2\sigma_0=(\tilde H
+iE\sigma_0)(\tilde H -iE\sigma_0)$ can be used to write 
\begin{eqnarray}
& G(iE) = \frac{i (iE\sigma_0 + H)}{2 E} & \left( (\tilde{H}-iE\sigma_0)^{-1}
\right. \nonumber \\
& & \left. - (\tilde{H}+iE\sigma_0)^{-1} \right)
\end{eqnarray}
Observe that both $H$ and $\tilde{H}$ appear in this expression, but 
$H$ only in the numerator. Since $(\tilde{H}\pm iE\sigma_0 )^{-1}$ is an 
analytic function of $\mu D$ in either the upper or lower $\mu D$--half plane
and $P(\mu D) = P(\mu)$ 
we can now perform the disorder integration.
The disorder averaged Matsubara Green function is translational invariant.
Performing a spatial Fourier transform we replace $-\nabla^2$ by $\xi = 
\epsilon_{\vec{k}} - \mu_0$ and find 
\begin{equation}
\langle G(iE)\rangle = - \frac{(iE +i\gamma)\sigma_0+ \xi\sigma_3 + 
{\hat \Delta}^{s \atop d}\sigma_1 }
{(E+\gamma)^2 + \xi^2 + ({\hat \Delta}^{s \atop d})^2} \equiv G(iE+i\gamma)\,.
\end{equation}
This is the Matsubara Green function of the pure system with the frequency
$iE$ shifted by the disorder parameter, $iE \rightarrow iE + i\gamma$.
It should be noted that for the local (isotropic) s--wave order parameter
discussed before the average over a Lorentzian distribution in Eq. 
\ref{gswave1} implies a shift $i\sqrt{\Delta^2+E^2}\to 
i\sqrt{\Delta^2+E^2}+i\gamma$.\\

To obtain the DOS for the d--wave case we approximate
the sum over the momenta $\vec{k}$ in 
standard fashion as $N_o \int_{-\infty}^{\infty} d\xi 
\int_0^{2\pi} {d\phi \over 2\pi}$ where $N_o$ is the density of states 
of the normal metal at the Fermi level, with 
the tetragonal Fermi surface approximated by a circle.
The result is 
\begin{equation}
N(E) 
=N_0  \int_0^{2\pi} \frac{d\phi}{2\pi} {\rm Im}\left( \frac{E + i\gamma}
{(\Delta_d^2(\phi) - (E +i \gamma)^2)^{1/2}}\right)
\end{equation}
where  the d--wave order parameter is approximated by $\Delta_d(\phi)= 
\Delta_d \cos (2\phi)$.
At $E=0$, $N(0) = N_0 \frac{2\gamma}{\pi \Delta_d} \mbox{ln}(4\Delta_d/\gamma)$
for $\gamma << \Delta_d$.
Thus, the DOS is nonzero at the
Fermi level for arbitrarily small disorder. For small $E$, $N(E)$
varies as  $E^2$. \\

For more general continuous distributions $P(\mu)d\mu$ the averaged DOS
can be estimated using again the analytic structure of ${\tilde G}$.
Applying the ideas of Ref. \cite{ziegler2}, one can derive a lower bound
by a decomposition of the
lattice into finite sub--squares. The average DOS on an isolated sub--squares
can
be estimated easily. Moreover, the contribution of the connection between the
sub--squares to the average DOS can also be estimated. A combination of both
contributions leads to $\langle N(0) \rangle\ge c_1\min_{-\mu_1\le\mu\le\mu_1}
P(\mu)$, 
where $c_1$ and $\mu_1$ are distribution dependent positive constants. 
In particular, $\mu_1$ 
must be chosen such that the spectrum of
$H(\mu_0=0)=-\nabla^2\sigma_3-i{\hat\Delta}^{s\atop d}D\sigma_1$ is
inside the interval $[-\mu_1,\mu_1]$. For all unbounded
distributions, like the Gaussian distribution 
used in Ref. \cite{Nersetal}, as well as compact distributions with
sufficiently large support this estimate
leads to a nonzero DOS at the Fermi level.\\

{\it Discussion and comparison to other methods.}
The major result in the d--wave (extended s--wave) case with Lorentzian
disorder is the presence of a finite purely imaginary self energy $\Sigma_0
= -i \gamma\sigma_0$ due to nonmagnetic disorder which leads to a {\it 
nonzero} DOS at the Fermi level. The latter is in 
qualitative agreement with standard theories based on the 
self--consistent t--matrix approximation \cite{ph1,srink}
as well as with exact 
diagonalization studies in 2D \cite{XiangWheatley}. In contrast to 
such theories our self energy has no dependence on 
${\hat \Delta}^{s \atop d}$, i.e. it is the same as in the normal state.
In Fig. 1 we show a comparison of the self energies of our theory 
and the limits of the t--matrix approximation.\\
\begin{figure}[p]
\leavevmode\centering\psfig{file=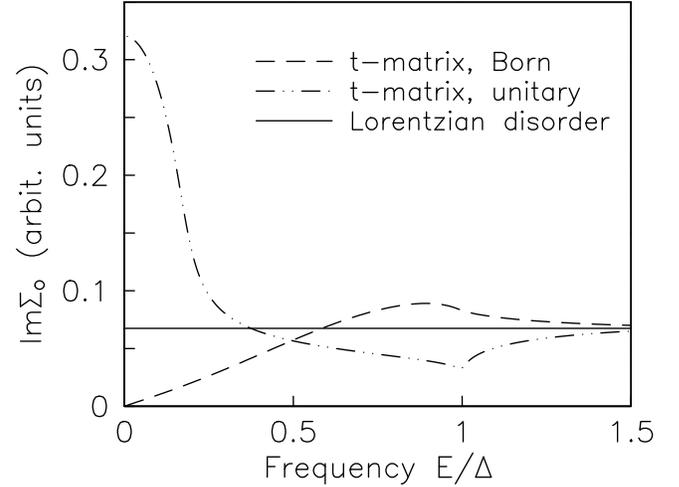,width=2.2 in}
\caption{Imaginary part of the self energy vs. frequency.
For Lorentzian disorder (solid line) the self energy is constant $i\gamma$.
The self energy of the self--consistent t--matrix
approximation in the unitary scattering limit (dashed--dotted line) 
behaves $\propto (\delta \Delta)^{1/2}$ at zero frequency. For Born scattering
(dashed line) the value at zero frequency is nonzero, but exponentially small.
We have adjusted
the impurity concentration to obtain equal normal state self energies
for the t--matrix results.}
\end{figure}
A drawback of the model with Lorentzian disorder
is that impurity concentration 
does not appear explicitly in the theory.  Whereas
in the t--matrix approach we have with the impurity concentration 
and the scattering
strength (or phase shift) two 
parameters associated with disorder, in the present model we have only 
$\gamma$, the  width of the Lorentzian. A way of 
making a connection is by comparing the variance of the 
Lorentzian  distribution ($\gamma$)
and the variance of the distribution underlying the t--matrix approximation,
which is a bimodal distribution  of a chemical potential $\mu = \mu_0$ with
probability $1-\delta$ ($\delta$ being the dimensionless impurity 
concentration) and $\mu=\mu_0 +V$ with probability $\delta$ ($V$ being
the scattering potential).
The variance ${\rm Var}_{\mu}$ of this distribution is determined by
\begin{equation}
{\rm Var}_{\mu}^2 = 
 \langle \mu^2\rangle  - \langle \mu\rangle^2 = V^2(\delta -\delta^2)\,\, .
\end{equation}
For small concentrations of impurities, $\delta << 1$,
we find ${\rm Var}_{\mu} = V \delta^{1/2}$. 
The $\delta^{1/2}$ behavior is also found for ${\rm Im} \Sigma_o (E=0)$
in the t-matrix approach for 
strong  scattering. Since in our model the variance of the distribution is 
also the imaginary part of the self energy, this suggests that
our model is closer to the strong scattering limit of the t--matrix 
approximation than the Born limit.\\

Finally, we comment on the discrepancies between our result and 
the calculation of 
Nersesyan et al., who found a power law for the averaged DOS with 
Gaussian disorder. 

One might question the analysis of Nersesyan et al.
because of the use of the replica trick, which is a dangerous
procedure in a number of models. \cite{ziegler3}  However, Mudry et al.
\cite{mudry} have obtained identical results for the {\it continuum}
problem of Dirac fermions in the presence of a random gauge field
using supersymmetry methods.  We therefore believe that the
crucial difference between our results and those of Ref. \cite{Nersetal}
occurs in the passage to the continuum and concomitant mapping of the
site disorder in the original problem onto the random gauge field.  Only
in the continuum case is there a direct analogy between disorder
in the chemical potential and a gauge field; on the lattice, 
gauge fields and chemical potential terms enter quite differently.  First,
chemical potential terms are local while gauge fields are defined on 
bonds.  Furthermore, chemical potential disorder enters linearly
in the Hamiltonian while gauge fields enter through the Peierls
prescription as a phase in the exponential multiplying the kinetic energy.\\

Disorder of the gauge field type is furthermore nongeneric 
even in the continuum,
as discussed by Mudry et al., who 
showed that the critical points of the
system with random gauge field are {\it unstable}
with respect to small perturbations by other types of disorder.\cite{foot1} 
We expect that a proper mapping of the lattice Dirac fermion 
or d-wave superconductor problems to  continuum
models will inevitably generate disorder other than random gauge fields.
Therefore, we believe that our result of a finite DOS at the Fermi level 
is the generic case for a d-wave superconductor 
in two dimensions.\\

{\it Conclusions.} We have computed the single particle Green function and DOS
for  a model of a superconductor with nonmagnetic impurities.
For an isotropic s--wave superconductor, we recover standard results; in
particular, Anderson's theorem is reproduced.
Our calculations for the disorder-averaged d- and extended s-wave
propagators show the  DOS is nonzero for
all energies, provided the
distribution of the chemical potential is continuous and of sufficient width.
The disorder average has been performed
exactly in the case of a Lorentzian distribution.
Our calculation casts doubt on
the result by Nersesyan et al.,
who found a power law for the averaged DOS with Gaussian disorder, and
suggests that the standard t--matrix approach to disordered d--wave
superconductors is qualitatively sufficient.\\

{\it Acknowledgements.}  The authors gratefully acknowledge
discussions with R. Fehrenbacher and C. Mudry. 
They are also grateful for the support of
the Institute for Fundamental Theory, University of Florida, and the
Institut f\"ur Theorie der Kondensierten Materie, Universit\"at 
Karlsruhe.\\


\begin{references}

\bibitem{AGL} For recent reviews see J. Annett, N. Goldenfeld, and 
A. Leggett, LANL cond-mat/9601060 and D.J. Scalapino, 
Physics Reports {\bf 250}, 329 (1995).


\bibitem{GorkovKalugin} L.P. Gor'kov and P.A. Kalugin, Pis'ma Zh. Eksp. Teor.
 Fiz. {\bf 41}, 208 (1985) [JETP Lett. {\bf 41}, 253 (1985)].

\bibitem{UedaRice} K. Ueda and T. M. Rice in {\it ``Theory of Heavy Fermions 
and Valence Fluctuations''}, eds. T. Kasuya and T. Saso, Springer Series
in Solid State Sciences, vol. 62, p. 267 (1985).

\bibitem{gross} F. Gross et al., Z. Phys. {\bf B 64}, 175 (1986).

\bibitem{Ishida} Y. Kitaoka et al., J. Phys. Soc. Japan {\bf 63}, 2052 (1994);
D. A. Bonn et al., Phys. Rev. {\bf B 50}, 4051 (1994);
J. Giapintzakis et al., Phys. Rev. B  ; preprint 1995.

\bibitem{HPSPRB}P.J. Hirschfeld, W.O. Putikka,
and D.J. Scalapino, Phys. Rev. {\bf B 50}, 10250 (1994).

\bibitem{ph1}
P.\,J.\,Hirschfeld et al.,
Sol.\, St.\, Commun. \, {\bf 59}, 111 (1986).

\bibitem{srink}
S.\,Schmitt-Rink et al., Phys.\, Rev.\, Lett. {\bf 57},
2575 (1986).

\bibitem{Nersetal} A. A. Nersesyan, A. M. Tsvelik and F. Wenger, Phys. Rev. 
Lett. {\bf 72}, 2628 (1994); Nucl. Phys. {\bf B 438}, 561 (1995).

\bibitem{MW} 
N. D. Mermin and H. Wagner, Phys. Rev. Lett. {\bf 17}, 1133 (1966).

\bibitem{Anderson} P. W. Anderson, J. Phys. Chem. Solids {\bf 11}, 26 (1959).

\bibitem{RL}  P. A. Lee and T. V. Ramakrishnan, Rev. Mod. Phys. {\bf 57}, 287 
(1985).

\bibitem{ziegler2}
K. Ziegler, Comm. Math. Phys. {\bf 120}, 177 (1988).

\bibitem{ziegler1}
K. Ziegler, Phys. Rev. {\bf B 53}, 9653 (1996).

\bibitem{Leedisorder} Y. Hatsugai and P.A. Lee, Phys. Rev. {\bf B 48}, 
4204 (1993).

\bibitem{XiangWheatley} T. Xiang and J.M. Wheatley, Phys. Rev. {\bf B 51}, 
11721 (1995).

\bibitem{ziegler3}
K. Ziegler, Phys. Rev. Lett. {\bf 73}, 3488 (1994).

\bibitem{mudry} C. Mudry, C. Chamon and X.-G. Wen, preprint, cond-mat/9509054

\bibitem{kogan} J.-S. Caux, I. I. Kogan and A. M. Tsvelik, preprint, 
hep-th/9511134



\bibitem{foot1} It should
be mentioned that even in the case of a random gauge field it is not
established
\cite{mudry,kogan} that the DOS indeed  vanishes at the Fermi level.

\end{references}
\end{document}